\documentclass[[aps,prb,reprint,superscriptaddress,floatfix]{revtex4-1}
\usepackage{graphicx}

\usepackage{amsmath}
\usepackage{amssymb}
\usepackage{hyperref}
\usepackage{bm}

\usepackage{xcolor}

\newcommand{\e}{{\bm e}}
\newcommand{\A}{{\bf A}}

\newcommand{\E}{{\bf E}}

\renewcommand{\a}{{\bf a}}
\newcommand{\n}{{\bf n}}
\newcommand{\np}{{n'}}

\renewcommand{\j}{{\bf j}}

\newcommand{\q}{{\bm{q}}}
\renewcommand{\r}{{\bm{r}}}

\DeclareMathOperator{\sign}{sign}
\DeclareMathOperator{\erf}{erf}
\DeclareMathOperator{\tr}{tr}

\begin{document}
	
\title{Unidirectional plasmonic edge modes on general two-dimensional materials}
\author{T. Stauber}
\affiliation{Materials Science Factory, Instituto  de Ciencias de Materiales de Madrid, CSIC, E-28049, Madrid, Spain}
\author{A. Nemilentsau}
\affiliation{Department of Electrical \& Computer Engineering, University of Minnesota, Minneapolis, Minnesota 55455, USA}
\author{T. Low}
\affiliation{Department of Electrical \& Computer Engineering, University of Minnesota, Minneapolis, Minnesota 55455, USA}
\author{G. G\'omez-Santos}
\affiliation{
Departamento de F\'{\i}sica de la Materia Condensada, Instituto Nicol\'as Cabrera and Condensed
	Matter Physics Center (IFIMAC), Universidad Aut\'onoma de Madrid, E-28049 Madrid, Spain}
 \date{\today}

\begin{abstract}
We investigate the field and spin-momentum coupling of edge plasmons hosted by general two-dimensional materials and identify sweet spots depending on the polarisation plane, ellipticity and the position of an electric dipole relative to the plane and edge. Exciting the dipole at these sweet spots by propagating light leads to uni-directional propagating edge plasmons or edge modes suppression. We also extend previous approximate treatments [A. Fetter Phys. Rev. B {\bf 32}, 7676 (1985)] to include anisotropy and hyperbolic systems, elucidating its predictions for the existence of edge modes. A thorough assessment of the approximate description  is carried out, comparing its spin-momentum coupling features in the near field with exact results from Wiener-Hopf techniques. Simulations are also performed confirming the overall picture.  Our results shed new light on the quest of chiral plasmonics in 2D materials and should be relevant for future experiments. 
\end{abstract}

\maketitle

\section{Introduction}
Chiral nano-photonics has attracted increasing attention especially in view of new routes to information technology \cite{Novotny12}, among which plasmonic circuitries have the advantage of a reduced length scale, compatible with nano-electronics \cite{Ozbay06}. Arguably, the most studied case of chiral plasmonics is the one of magneto-plasmons \cite{Fetter85,Volkov88,Mikhailov95,Wang11,Yan12,Crassee12,Kumada13,Kumada14,Cohen18}, leading to infrared topological plasmons in periodically patterned monolayer graphene \cite{Jin17} and superlattices \cite{Pan17}. Non-reciprocal plasmons can further be obtained by an external source-drain bias \cite{Behnaam15,Sabbaghi18,Bliokh18}, leading to collimated plasmon beam in the case of large drift current \cite{Duppen16}. And a non-trivial Berry curvature of the band-structure can also lead to chiral propagation of the edge mode by appropriately breaking time-reversal symmetry via, e.g., circularly polarized light \cite{Kumar16,Son16}. Exciting chiral (edge) plasmons at fixed frequency and momentum via a grating geometry would then lead to uni-directional (edge) plasmon propagation. Nevertheless, the opposite plasmons dispersions hardly differ from each other such that both propagation directions are usually excited. Also, plasmons are mostly excited by illuminating a point-dipole via propagating light that excites plasmons of all wave numbers and will thus also launch plasmons in the opposite direction. 

Here, we will investigate a more fundamental route to chiral nano-plasmonics based on the inherent spin-momentum coupling \cite{Lee12,Lin13,Bliokh15a,VanMechelen16,Lodahl17}, which (in some context) can be linked to the quantum spin Hall effect of light \cite{Bliokh15b}. From Maxwell's equations it follows that for strongly confined plasmons, the electric field is either left or right circularly polarized depending on the direction of propagation. Providing a one-dimensional wave guide, circularly polarized light at oblique incidence can then only excite plasmons that propagate in one direction according to the spin-momentum coupling \cite{Lee12}. Anisotropic materials such as black phosphorous\cite{Low14,Goncalves17} or hyperbolic materials\cite{Ma18} could thus provide a intrinsic propagation direction leading to plasmons with a high collimation of electric field energy \cite{Nemilentsau19}. The concept can also be extended to Janus and Huygens dipoles \cite{Picari18} and hyperbolic materials.

Nowadays, the most prominent 2D plasmonic material is graphene, mainly due to its high intrinsic mobility/life-times and easy tuneability of its plasmonic resonances \cite{Koppens11,Grigorenko12,Stauber14,Peres16,Basov16,Low17}. They can be excited and detected by means of infrared nano imaging \cite{Fei12,Chen12}. Since graphene's plasmons are strongly confined, a left circularly polarized dipole in the $xz$-plane on top of graphene placed at $z=0$ will excite plasmons propagating only in the positive $x$-direction and not in the opposite, negative $x$-direction. But graphene is an isotropic material and the circularly polarized dipole in the $xz$-plane is elliptically polarized in all other directions. Since elliptically polarized light can be decomposed into a left and right circularly polarized component, a dipole usually excites plasmons that propagate in both, forward and backward direction, albeit with different intensities. This naturally limits the usefulness of the spin-momentum locking in terms of collimation of the plasmonic energy current in only one direction.

We will, therefore, mainly consider plasmonic edge-modes of an isotropic medium such as graphene since they provide a natural wave guide at the border of the sample \cite{Nikitin11}. Recently, they have been revealed by infrared nano imaging, i.e., scattering-type scanning near-field optical microscopy (s-SNOM) \cite{Fei15,Nikitin16}. It is generally argued that graphene edges should be separated by an effective working distance to avoid the overlapping of localized plasmon modes with the bulk modes, important for the design of graphene-based plasmonic circuits and devices.\cite{Qingyang17} But here, we will focus on the edge modes as principle players of plasmonic circuits similar to the proposal of Refs. \cite{Pan17,Bisharat17}. In fact, our results will in general be valid for any two-dimensional system with an arbitrary conductivity tensor. For hyperbolic materials,\cite{Ma18} i.e., for systems with a conducting and non-conduction principle axis, we can rule out the existence of edge plasmons within the Fetter approximation.\cite{Fetter85}

The paper is organised as follows. In Section \ref{SimpleGeometries}, we discuss the spin-momentum coupling for simple geometries modelling bulk and edge plasmonic modes. In Section \ref{EdgeMode}, the spin-momentum coupling for a two-dimensional plasmonic edge mode is discussed analytically within the Fetter\cite{Fetter85} approximation, extended to cover general elliptic and hyperbolic cases, studying the condition for the emergence of edge states. We also compare the approximate results  for the near field with the exact Wiener-Hopf\cite{Volkov88} treatment. In Section \ref{Simulations}, we then verify our conclusions by COMSOL simulations launching unidirectional plasmonic edge modes by circularly polarized dipoles.  We close with conclusions and summary and add two appendices providing details of the analytical treatment. 

\section{Spin-momentum coupling of simple geometries}
\label{SimpleGeometries}
The spin-momentum coupling is a consequence of the transverse nature of a confined gauge-field. But it can also be obtained from the static Poisson equation. In the following, we will discuss simple geometries related to bulk and edge plasmons and introduce basic concepts such as the ellipticity. 
\subsection{Plane interface}
For an interface at $z=0$ and choosing in-plane momentum $\q=\tau q\e_x$ with $q>0$, unit vector along the $x$ axis $\e_x $ and propagation sense $\tau=\pm$, the vector potential in half-space $z<0$ ($m$=1) and half-space $z>0$ ($m$=2) can be written as
\begin{align}
\A_m(x,z)=A_0\sum_\q e^{i\tau qx-q_m'|z|} (\e_x+sgn(z)\tau i\frac{q}{q_m'} \e_z)\;,
\end{align}
where $q_m'=\sqrt{q^2-\epsilon_m(\omega/c)^2}$ is the transverse wave number, $\epsilon_m$ the dielectric constants of medium $m$, and $\e_z $ the unit vector along the $z$ axis. 

Due to the phase factor $i\frac{q}{q_m'}$, there is a rotation of the polarisation vector in the $xz$-plane and the sense of rotation depends on the propagation direction. We can define the ellipticity of the electric fields $\E=-\partial_t \A$ as the ratio between the field perpendicular and parallel to the propagation divided by a phase $i$:
\begin{align}
\psi=\frac{E_\perp}{iE_\parallel}=\frac{E_z}{iE_x}=sgn(z)\tau\frac{q}{q_m'}
\end{align}

Generally, the polarisation is elliptically polarized. In the case of graphene, however, the plasmons are strongly confined, such that $q'\approx q$ and therefore $\psi\approx\pm1$. This makes graphene's plasmon mode {\it circularly} polarized. Also note that for bulk modes, the polarisation plane is always normal to the interface.

\subsection{Infinite line charge}\label{linecharge}
Let us consider the potential $\Phi $ of an infinitely long line charge along the $x$ axis described by the following Poisson equation:
\begin{align}
\label{Poisson}
-{\bm \nabla}^2\Phi(\r,x)=\rho(x)\delta(\r)
,\end{align}
where $\r$ represents the perpendicular vector position.
This could be a very elementary model for a  plasmonic edge mode  of a half-infinite plane as in Fig. \ref{halfplane}, if we take a propagating wave in $x$-direction with wave number $q$, i.e., $\rho(x)=e^{i\tau qx}$. Writing $\Phi(\r,x)=e^{i\tau qx}g(\r) $,
Eq. (\ref{Poisson}) then reduces to the problem of finding the Green's function $g(\r)$ of a 2D massive particle:
\begin{align}
g(\r)=\frac{1}{2\pi}\int_0^\infty dk\frac{kJ_0(kr)}{k^2+q^2}=\frac{1}{2\pi}K_0(qr)
\end{align}
with Bessel function of first kind $J_0$, and  modified Bessel function of second kind $K_0$. This model can be extended to a charged cylinder of radius $a$ without major changes. 

Again, we are interested in the ratio of the absolute value of the electric fields with respect to the perpendicular and parallel direction with respect to the propagation:
\begin{align}
\label{PsiLineCharge}
\psi=\frac{E_\perp}{iE_\parallel}=\frac{E_r}{iE_x}=\tau\frac{K_0'(qr)}{K_0(qr)}
,\end{align}
where $K_0' $ means derivative. Note that this ratio   depends on the radius $r$, and on the propagation sense $\tau$.  For $r\gg1/q$, this field becomes circularly polarized.  Further, the polarisation plane is always that containing the line charge and the observation point. 

 This analysis suggests that the field of an edge mode {\it outside} the 2D material should be nearly circularly polarized and the polarisation plane is rotating around the wire. In the following section, we will discuss how this picture is modified when the extended, half-infinite,  2D material plane is taken into account.

\section{Spin-momentum coupling of two-dimensional plasmonic edge modes}
\label{EdgeMode}
\begin{figure}
	\includegraphics[width=1.1\columnwidth]{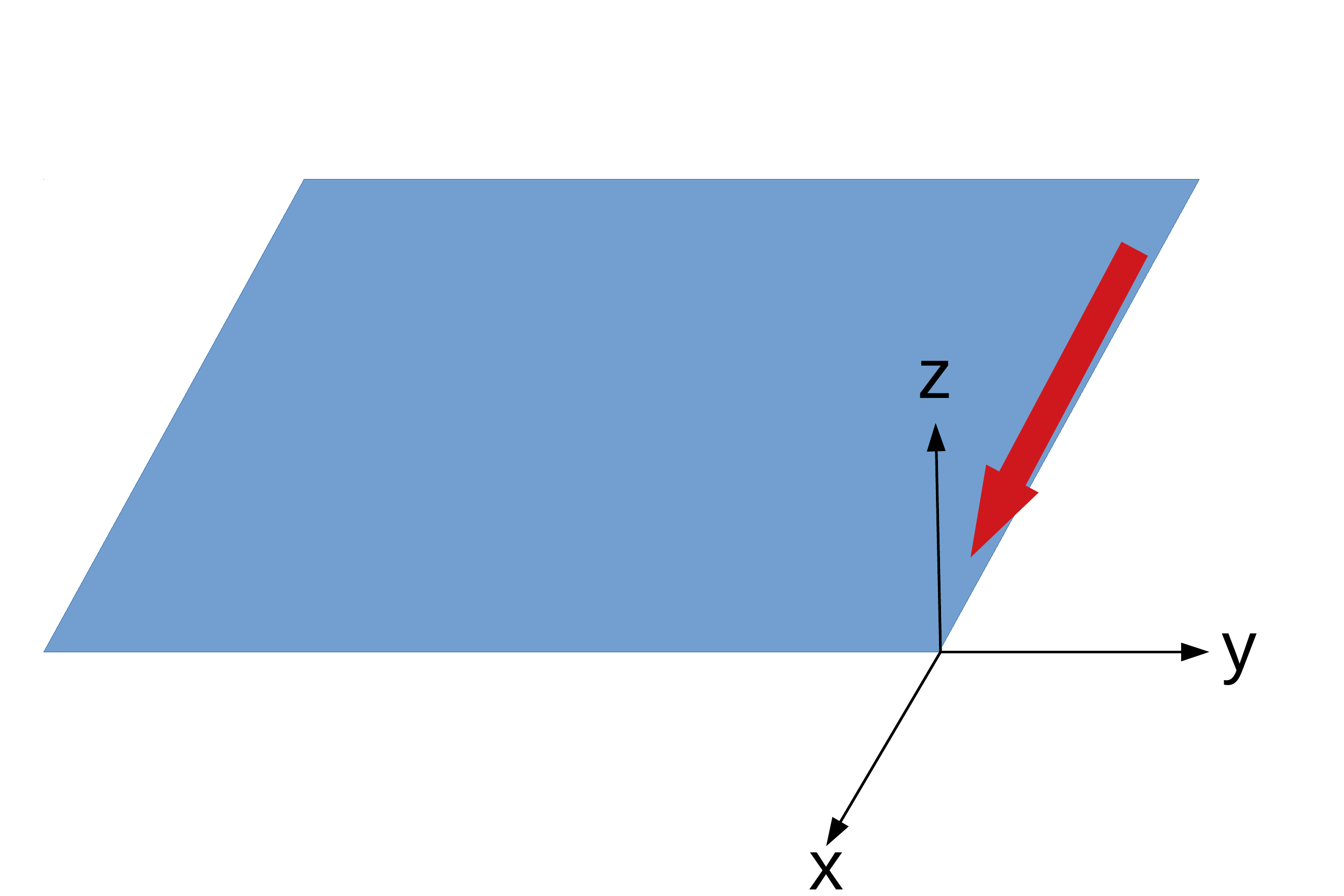}
 	\caption{ Geometry of the material half-plane located at $z=0$  and occupying the region $y<0$, so the edge coincides with the $x$ axis. The thick (red) arrow illustrates an edge plasmon propagating along the positive $(\tau=+1)$ $x$ direction.
 \label{halfplane}}
\end{figure}

Charged graphene, as any two-dimensional electron gas, hosts bulk density fluctuation excitations called plasmons. In addition, edge localized plasmons emerge when the geometry is restricted to a half-infinite plane. We choose the geometry depicted in Fig. \ref{halfplane}, where the material half-plane is located at $z=0$ and occupies the region $y<0$, so the edge coincides with the $x$ axis.
Although a  complete characterization of this edge mode in the static limit can be achieved by means of Wiener-Hopf\cite{Volkov88} techniques, here we follow the simplified approach pioneered by Fetter\cite{Fetter85,Wang11,Song16} to study its near field.

We are interested in the near fields produced by self-sustained currents inside a 2D conducting plane $\j=-\chi\A$ with local current-current response depending on the (say) $y$-component:
\begin{align}
\label{BoundaryChi}
\chi(y)=\chi f(y)\;,\;\rm{with}\;\chi=\left(
\begin{array}{cc}
\chi_{xx}&\chi_{xy}\\
\chi_{yx}&\chi_{yy}
\end{array}\right)\;,
\end{align}
where $f(y)$ denotes an arbitrary function. Note that we also introduce an arbitrary response matrix that could describe any possible physical system in 2D including hyperbolic materials. The response matrix $ \chi$, related to the conductivity by $\chi = -i \omega \sigma $,  can be termed  the Drude matrix, as it reduces to the Drude weight in the isotropic case.
The bulk plasmons propagating in $x$-direction between two dielectrics $\epsilon_1$ and $\epsilon_2$ have the dispersion $\omega_b^2= \tfrac{\chi_{xx}q}{2 \varepsilon_0\epsilon}$ with $\epsilon=(\epsilon_1+\epsilon_2)/2$.

We will solve the Maxwell equations with the constitutive equation of Eq. (\ref{BoundaryChi}). In the Appendix A , we outline how retardation effects can be included in the basic equations characterised by the generating function 
\begin{align}
L_m^q(y,z)=\frac{1}{2}\int_{-\infty}^\infty \frac{dq_y}{2\pi}e^{iq_yy}e^{-q_m|z|}\frac{q_1'q_2'}{(q_1')^2q_2'+(q_2')^2q_1'}\;.
\end{align}
With this, we can relate the vector potential  $\A_m(\r)=A_0\sum_qe^{i\tau qx}\a_m(q,y,z)$ to the current:
\begin{align}
\label{EdgeGaugeField}
\a_m(q,y,z)=&-\frac{\mu_0}{\Omega^2}\left(
\begin{array}{cc}
q^2-\Omega^2&-iq\partial_y\\
-iq\partial_y&-\partial_y^2 -\Omega^2
\end{array}\right)\notag\\
\times&\int_{-\infty}^\infty dy'  L_m^q(y-y',z)\j(y')\;,
\end{align}
with $\Omega^2=\frac{\epsilon_1q_2'+\epsilon_2q_1'}{q_1'+q_2'}\frac{\omega^2}{c^2}$. We note that this equation is including retardation which has not been discussed so far. Retardation effects are important in the context of chirality as introduced by Tang and Cohen.\cite{Tang10,Tang11}
\begin{figure}
	\includegraphics[width=0.99\columnwidth]{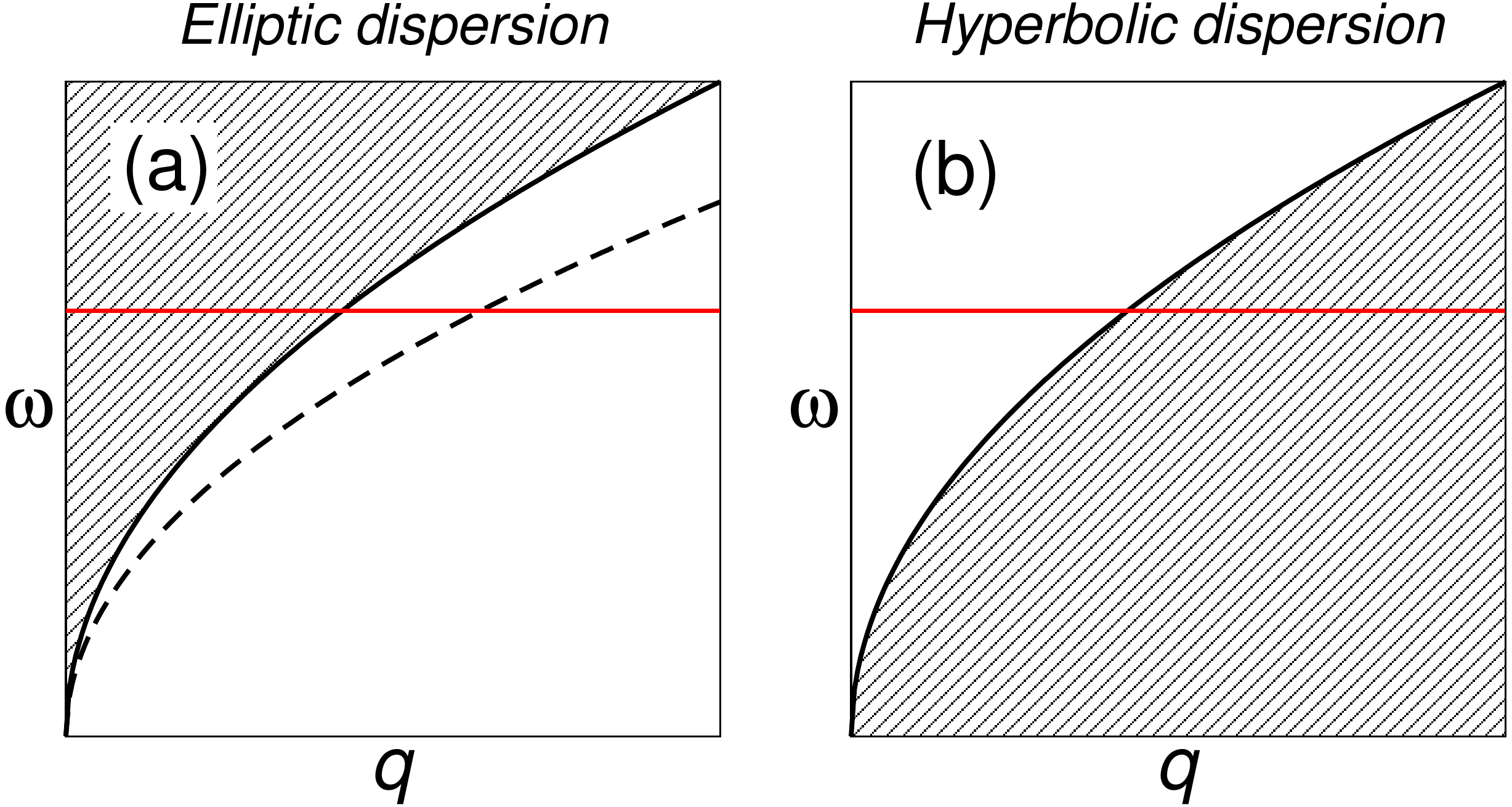}
 	\caption{(a) Schematic illustration of plasmons near an edge as a function of wavevector along the edge for an elliptic (ordinary) material. Dashed line: edge plasmon. The shaded area represents the bulk plasmon continuum.
(b) As in the (a) panel, but for a hyperbolic material. No edge plasmon detaches from the continuum in this case (see text). Both plots assume a frequency independent Drude matrix, taken to be the true Drude matrix at, for instance,  the frequency indicated by the red line: the nominal line of validity.
\label{figedgedispersion}}
\end{figure}

\subsection{Analytical approximation}
To proceed analytically, we will set $f(y)=\Theta(-y)$, neglect retardation effects and approximate the generating function as:
\begin{align}
\label{Green}
L_m^q(y,z=0)\approx \frac{1}{\sqrt{8}}e^{-\sqrt{2}q|y|}
\end{align}
a procedure first proposed by Fetter\cite{Fetter85} that amounts to replacing the non-local Kernel of Eq. \ref{EdgeGaugeField} by one amenable to solution as a differential equation for the $y$ variable in the sheet plane.
We then obtain $\a_m(q,\pm|y|,z)=\a_{m,\pm}e^{-\kappa_\pm|y|}e^{-\kappa_z^\pm|z|}$:
\begin{align}
\label{GaugeField}
\a_{m,\pm}=\left(
\begin{array}{c}
\tau\\
\pm i\tilde\kappa_\pm\\
(-1)^mi\tilde\kappa_z^\pm
\end{array}\right)
\end{align}
where we have introduced the  dimensionless quantities $\tilde\kappa_\pm=\kappa_\pm/q$ and $\tilde\chi_{n,n'}=\frac{q}{\epsilon_0\epsilon\omega^2}\chi_{n,n'}$, with $\tilde\kappa_{+}=\sqrt{2}$. Defining to facilitate the notation in the material region $\tilde\kappa_y=\tilde\kappa_-$, then the putative edge mode emerges as the solution of the following two equations:
\begin{align}
\label{kappaY}
\tilde\kappa_y\equiv\tilde\kappa_-= & \frac{i\tau\tilde\chi_{yx}-\sqrt{2}}{1-\tilde\chi_{yy}} \;,\\
\label{DispersionCondition}
2-\tilde\kappa_y^2= &\tilde\chi_{xx}-\tilde\kappa_y^2\tilde\chi_{yy}-i\tau(\tilde\chi_{xy}+\tilde\chi_{yx})\tilde\kappa_y\;,
\end{align}
as shown in the Appendix A.
 
\subsection{Solutions}
The solutions to the edge mode equations depend on the material Drude matrix, itself a function of frequency, a fact that apparently precludes the possibility of saying something general. Nevertheless, one can make progress by pretending that the Drude matrix at the target frequency remains constant for all frequencies. In this way, one can attain rather general conclusions on the existence of edge modes that are valid, at least, on the chosen $\omega=\text{constant}$ line in phase space, for instance,  the horizontal lines of Fig. \ref{figedgedispersion}. The conclusions of such analysis for different material classes are described in what follows. 

\subsubsection{Elliptic (or ordinary) systems}
We assume a non-absorbing material with, for the moment, time-reversal symmetry (TRS). The Drude matrix of Eq. \ref{BoundaryChi} is then made of real entries with $\chi_{xy} =\chi_{yx}$ owing to TRS. The ordinary nature of the material is taken to mean a positive determinant for $\chi$:  
\begin{equation}
\label{DetOrdinary}
\det (\chi) = \chi_{xx} \chi_{yy} - \chi_{xy}\chi_{yx} > 0
,\end{equation}
If, in addition, one assumes $\chi$ to be positive definite,  a rotation to principal axes produces two positive eigenvalues which, in the isotropic case, reduces the matrix to a scalar, the Drude weight  of an ordinary metal. 
Such system supports bulk plasmons with {\it elliptic} dispersion given by $\omega^2_b(q)=\frac{1}{2\epsilon_0\epsilon q} q_n\chi_{n,n'}q_n' $, leading in the isotropic case to the well-known expression $\omega^2_b(q)=\frac{D q}{2\epsilon_0\epsilon }$, where $q$ is here the bulk wavevector $q=\sqrt{q_x^2+ q_y^2}$, and $D$, the Drude weight.\cite{Stauber14}

As shown in the Appendix B,  an edge state below the continuum of bulk plasmons for systems satisfying Eq. \ref{DetOrdinary} will exist whenever
\begin{equation}
\label{OrdinaryCondition}
\chi_{yy} > 0\;,
\end{equation}
which together with Eq. \ref{DetOrdinary}, amounts to a positive definite Drude matrix $\chi$, i.e., edge plamons will emerge whenever the bulk supports elliptic ({\it ordinary} or {\it metallic}) plasmons. This situation is schematically depicted on the (a) panel of Fig. \ref{figedgedispersion}. In the isotropic case considered by Fetter\cite{Fetter85}, the edge mode is given by $\omega_e^F(q)= \sqrt{\tfrac{2}{3}}  \,\omega_b(q)$, reasonably close to the exact solution of Volkov and Mikhailov\cite{Volkov88} $\omega_e^{VM}(q) \approx 0.9  \,\omega_b(q)$. Therefore, our results generalize to the anisotropic case the existence of edge modes in ordinary systems.   

In the absence of TRS such as, for instance, with a  perpendicular magnetic field present, the off-diagonal entries of $\chi$ can acquire an imaginary component, $\chi_{xy} =  \chi'_{xy} + i \chi_c $, and $\chi_{yx} =  \chi'_{yx} - i \chi_c $, where the real part is again symmetric $\chi'_{xy}=\chi'_{yx} $, and the imagnary part, $\chi_c, $ would be proportional to the cyclotron frequency. In that case,  the analysis of Appendix B predicts, to lowest order in $\chi_c$, a linear splitting of the edge mode depending on the propagation direction, just as in the isotropic case\cite{Fetter85,Volkov88}.

Quite generally, if an edge plasmon exists, it is shown in the Appendix B that it satisfies the following dispersion relation within the present approximation:
\begin{equation}
\label{FetterDispersion}
\omega_e^F(q)= \sqrt{\frac{\chi_{xx} \chi_{yy} - \chi'^{\,2}_{xy} - \chi_c^2}{2 \chi_{yy} +\chi_{xx} -2\sqrt{2} \tau \chi_c} \;  \frac{q}{\epsilon_0\epsilon}  }
\end{equation}

\subsubsection{Hyperbolic systems}
The Drude matrix of a hyperbolic, non-absorbing material with TRS is again real and symmetric, but now with a  negative determinant:  
\begin{equation}
\label{DetHyperbolic}
\det (\chi) = \chi_{xx} \chi_{yy} - \chi_{xy} \chi_{yx} < 0
.\end{equation}
Such system also supports bulk plasmons with dispersion given by $\omega^2_b(q)=\frac{1}{2\epsilon_0\epsilon q} (q^2_+ \chi_+ + q^2_- \chi_-)$, where $q_{\pm}$ are the wavevector components along principal axes,  with corresponding Drude entries now  of opposite sign: $\chi_+ > 0 $ and $ \chi_- < 0$, therefore, a {\it hyperbolic} dispersion. Edge states, if present, would emerge above the bulk continuum, as illustrated in the (b) panel of Fig. \ref{figedgedispersion}. The analysis of the Appendix B shows that no such localized edge states exist   in the  hyperbolic case, irrespective of the degree of anisotropy.

\subsection{Validity of the approximation}
Let us estimate the quality of the Fetter approximation with respect to the boundary condition $\epsilon_2E_z(z=0^+)-\epsilon_1E_z(z=0^-)=-\omega\epsilon2\tau\kappa_y=\rho/\epsilon_0$. For $y>0$, we have $\rho=0$, and the boundary condition is trivially satisfied. But for $z\to\infty$, we would expect a decay of the vector potential. Also, there should not be any discontinuity of the vector potential  at $y=0$ for large $z$. Our solution can thus only be valid close to the edge and this restriction should be due to the approximation of Eq. (\ref{Green}). 

For $y<0$, the boundary condition would imply $2\tilde\kappa_y=2-\tilde\kappa_y^2$ which is only fulfilled for $\tilde\kappa_y=\sqrt{3}-1$. We will thus only expect a solution to well describe the edge mode for decay length with $2/3\sim\tilde\kappa_y\sim4/5$, allowing a relative error of up to $10$\%.

We can proceed with an analytical discussion by noting that the  in-plane potential of the edge-mode penetrates exponentially in the bulk, a feature expected to hold true also for the exact solution. This fact can be understood as an imaginary component for the $y$ component of momentum, $q_y= -i \kappa_y$. Making use of the sheet response and the continuity equation, the frequency of the edge plasmon must fulfill 
\begin{align}
2\sqrt{1-\tilde\kappa_y^2}=\tilde\chi_{xx}-\tilde\kappa_y^2\tilde\chi_{yy}-i\tau(\tilde\chi_{xy}+\tilde\chi_{yx})\tilde\kappa_y\;.
\end{align}
Comparing this with the dispersion obtained from the Fetter approximation, Eq. (\ref{DispersionCondition}), we see that it is only consistent up to order $\mathcal{O}(\tilde\kappa_y^4)$. The approximated dispersion is thus always red-shifted and for $2/3\sim\tilde\kappa_y\sim4/5$, we have relative errors of $4-12$\%. This analysis can be verified in the case of an isotropic response matrix where the exact solution of Volkov and Mikhailov gives $\omega_e^{VM}(q) \approx 0.9  \,\omega_b(q)$ and the Fetter approximation $\tilde\kappa_y=1/\sqrt{2}$ and $\omega_e^F(q)= \sqrt{\tfrac{2}{3}}  \,\omega_b(q)$, i.e., $\omega_e^{F}/\omega_e^{VM}=0.9$.  

All results hold for edge modes in an arbitrary system as long as they exist. For the special case of an isotropic system (with or without breaking time-reversal symmetry), we can further relate the exact edge mode energy to the exact bulk  mode energy as $\omega_e(q) = \omega_b(\kappa_z^-)$. This result will be useful for the next subsection where we discuss the ellipticity and polarisation plane of the edge mode, and  present a better assessment of the approximations.

\subsection{Polarization plane and ellipticity}
We are now in the position to discuss the polarisation plane and ellipticity for the edge mode of ordinary materials. Now, we have a field vector in all three directions and we have to group together the field components that have the same phase. 

From Eq. (\ref{GaugeField}) and for $\tilde\kappa_\pm<1$, Maxwell's equations dictate $\tilde\kappa_z^\pm=\sqrt{1-\tilde\kappa_\pm^2}$; for $\tilde\kappa_\pm>1$, this would result in an imaginary decay length and we could take the average of the two branches and set $\tilde\kappa_z^\pm=0$. We would then have the following result for the ellipticity in the Fetter approximation, given for the geometry of Fig. \ref{halfplane} by $\psi_-^{F}=E_{yz}/i E_x $, where $E_{yz}$ and $E_{x}$  are the $yz$-plane and $x$-axis field components:
\begin{align}
\psi_-^{F}=sgn(z)\tau\;,\;\psi_+^{F}=-\tau\sqrt{2}\;.
\end{align}
The field is thus always circularly polarized for $y<0$ and elliptically polarized for $y>0$. Notice that the mentioned inconsistencies in the field are to be expected because the approximation replaces the exact Maxwell kernel for an approximate one.

A fairer assessment is provided by regarding the whole Fetter procedure as a means to get an approximate surface charge density for the edge mode. This approximate surface density profile is taken to be
\begin{equation}\label{nF}
n_F(x,y)=  \rho_F(y) \frac{e^{i \tau q x}}{\sqrt{2 \pi}}
,\end{equation}
with $\rho(y)$ given up to a global scale by
\begin{equation}\label{rhoF}
  \rho_F(y) =  |q| \; e^{|q y|/\sqrt{2}} \;\Theta(-y) + \sqrt{2} \; \delta(y)
,\end{equation}
for the edge mode of the isotropic case, for simplicity. 
One can then obtain the near field associated with such charge using the exact Coulomb Kernel. The results of this calculation for the charge density and ellipticity in  the isotropic case are shown in the (a) and (b) panels of Figs. \ref{figAbsE}, \ref{figEllip} and \ref{figcombo}. They provide a rather satisfactory description of the near field of the edge mode when compared with the exact results, later obtained with far more involved Wiener-Hopt techniques.

A peculiarity of the near-field is that right on top and below the conducting sheet with isotropic response where $\tilde\kappa_y=1/\sqrt{2}$, the  field orientation in the $yz$ plane, $\theta= \arctan(E_z/E_{y})$,  reaches the asymptotic values of $\pm 135^{\circ}$, see (a) panels of Figs.  \ref{figEllip} and \ref{figcombo}. Being the charge localized near the edge, a naive expectation would suggest  a field oriented along the $-\e_y$ direction deep into the graphene sheet, as for the line charge of section \ref{linecharge}. This expectation is ill founded, as can be shown with the following exact reasoning. From the general discussion of the exponential decay of the edge mode in the bulk, we obtain the following relation for the asymptotic orientation of the near-field just above and below the graphene sheet:
\begin{align}\label{theta}
\tan(\theta) = \frac{E_z}{E_{y}} =-sgn(z)\frac{\kappa_z }{\kappa_y}=-sgn(z)
\frac{\tilde \kappa_z}{\sqrt{1-\tilde \kappa_z^2}}\;.
\end{align}
Let us now use the relation $\omega_e(q) = \omega_b(\kappa_z)$ valid for an isotropic system with or without a magnetic field. This leads to
\begin{align}\label{ratio}
\tilde \kappa_z= \left[ \frac{\omega_e(q)}{\omega_b(q)} \right]^2.
\end{align}
We believe this relation to hold true for the exact solution in the instantaneous approximation, as it only relies on the asymptotic exponential penetration of the edge excitation in bulk graphene, see discussion above. The Fetter solution only approximately fulfills this relation, owing to the lack of exact Coulomb self-consistency. Nevertheless, our value of $\theta= \pm 135 ^{\circ}$ in the case of a isotropic response compares favorably with $\theta\pm\approx 125^{\circ}$, a result obtained from  Eq. (\ref{theta})  for the exact\cite{Volkov88}  $\omega_e(q) \approx 0.9 \, \omega_b(q) $.

\subsection{Wiener-Hopf (exact) solution}
In this section we obtain the edge mode density and associated near field for an ordinary,  non-absorbing, isotropic material using the Wiener-Hopf technique developed in Ref. \cite{Volkov88}. The treatment provides the exact solution, at least within the certainly valid instantaneous approximation. The comparison with the previous approximate calculation confirms the latter to be a very good approach. 

We assume a half-infinite sheet in the plane $z=0$ with $y<0$ as in Fig. \ref{halfplane}. The surface charge density for a propagating  edge mode can be written as
\begin{equation}\label{nxy}
n(x,y)=  \rho(\tilde{y}) \frac{e^{i q x}}{\sqrt{2 \pi}}
,\end{equation}
where $\tilde{y}= |q| y $. The density $\rho(\tilde{y})$ can be obtained from its Fourier transform, $\tilde{\rho}(\tilde{k})$, which the Wiener-Hopf techniques  of Ref. \cite{Volkov88} allow to write as 
\begin{equation}\label{rho}
\tilde{\rho}(\tilde{k})= \tilde{\rho}_o f(\tilde{k})
,\end{equation}
with $f(\tilde{k})$ given by
\begin{equation}\label{fk}
f(\tilde{k}) = \exp\left[-\frac{X_+(\tilde{k})}{2 \pi i}\right]
,\end{equation}
where 
\begin{multline}\label{X+}
X_+(\tilde{k}) = \\
\int_{-\infty}^{+\infty}  d k' \left[\frac{1}{k'-\tilde{k}-i 0^+}- \frac{1}{k'-i 0^+}\right] \ln\left[\tilde{\kappa}^{-1}_z \sqrt{1+k'^2} -1\right]
,\end{multline}
with $ \tilde{\kappa}^{-1}_z=(\tfrac{\omega_b(q)}{\omega_e(q)} )^2 =1.2183$, under the assumption of a perfect Drude metal without dissipation. $f(\tilde{k})$ can be obtained {\it almost} analytically as 
\begin{equation}\label{fkanal}
f(\tilde{k}) = \frac{\sqrt{1-i \tilde{k}}}{1-i \tilde{k}/	\tilde{\kappa}_y} g(\tilde{k})
,\end{equation}
where $\tilde{\kappa}_y = \sqrt{1-\tilde{\kappa}^{2}_z}$, and the function $g(\tilde{k})$ is given by
\begin{equation}\label{gk}
g(\tilde{k}) = \exp \left[-\frac{\tilde{k}}{i \pi} \int_1^{\infty} dx \frac{\pi/2 - \arctan(\tilde{\kappa}^{-1}_z\sqrt{x^2-1})}{x (x-i \tilde{k} )}\right]
.\end{equation}
It is interesting to notice that 
$g(\tilde{k})$ is a smooth function and, if one ignores its $\tilde{k}-$dependence as  $g(\tilde{k})\approx 1 $,  a simple analytical expression for the density can be obtained, with correct functional behavior in the limits $y\to 0 $ and $y\to -\infty $. It is given by 
\begin{multline}\label{rhoanal}
\rho_{approx}= \sqrt{2 \pi} 	\tilde{\kappa}_y \sqrt{1-\tilde{\kappa}_y} \;e^{\tilde{\kappa}_y \tilde{y}} \; \erf(\sqrt{(1-\tilde{\kappa}_y)|\tilde{y}|})\;\Theta(-\tilde{y}) \\
+ \sqrt{2} \tilde{\kappa}_y \frac{e^{\tilde{y}}}{\sqrt{|\tilde{y}|}} \;\Theta(-\tilde{y})
,\end{multline}
where $\erf$ and $\Theta$ stand for error and Heaviside functions, respectively.

In any case, the results to follow use the (numerically) {\em exact} density. Writing the electric field as 
\begin{equation}
{\bf \mathcal{E}}(x,y,z)=\frac{e^{i q x}}{\sqrt{2 \pi}} \bm E(\tilde{y},\tilde{z})
,\end{equation}
the numerically obtained, exact Fourier transform of Eq. (\ref{rho})  allows its calculation by means of the following explicit formulas
\begin{multline}
E_z(\tilde{y},\tilde{z})= \frac{\tilde{\rho}_o}{2 \varepsilon_0} \sign(z) \times \\
\int_{-\infty}^{+\infty}  \frac{d\tilde{k}}{\sqrt{2 \pi}} \, f(\tilde{k}) \, e^{i\tilde{k}\tilde{y}} \,
e^{-\sqrt{1+\tilde{k}^2} \, |\tilde{z}|}
,\end{multline} 
\begin{multline}
E_y(\tilde{y},\tilde{z})= \frac{\tilde{\rho}_o}{2 \varepsilon_0}   \times \\
\int_{-\infty}^{+\infty}  \frac{d\tilde{k}}{\sqrt{2 \pi}} \, \frac{-i \tilde{k} f(\tilde{k})}{\sqrt{1+\tilde{k}^2}} \, e^{i\tilde{k}\tilde{y}} \,
e^{-\sqrt{1+\tilde{k}^2} \, |\tilde{z}|}
,\end{multline} 
and
\begin{multline}
E_x(\tilde{y},\tilde{z})= \frac{\tilde{\rho}_o}{2 \varepsilon_0} \sign(q)  \times \\
\int_{-\infty}^{+\infty}  \frac{d\tilde{k}}{\sqrt{2 \pi}} \, \frac{-i f(\tilde{k})}{\sqrt{1+\tilde{k}^2}} \, e^{i\tilde{k}\tilde{y}} \,
e^{-\sqrt{1+\tilde{k}^2} \, |\tilde{z}|}
,\end{multline} 
with $\tilde{z}= |q| z$.

\begin{figure}
	\includegraphics[width=0.99\columnwidth]{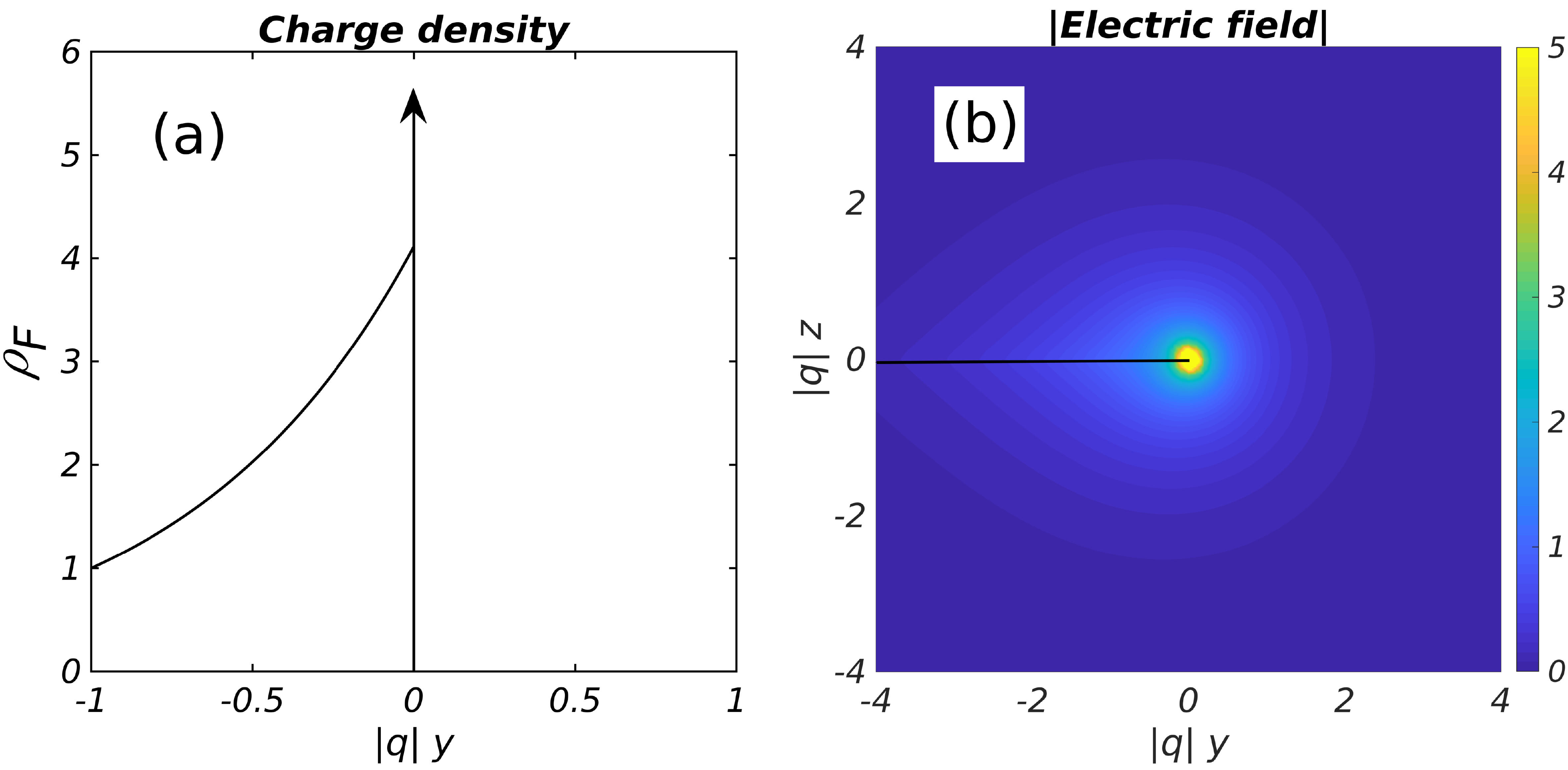}
        \includegraphics[width=0.99\columnwidth]{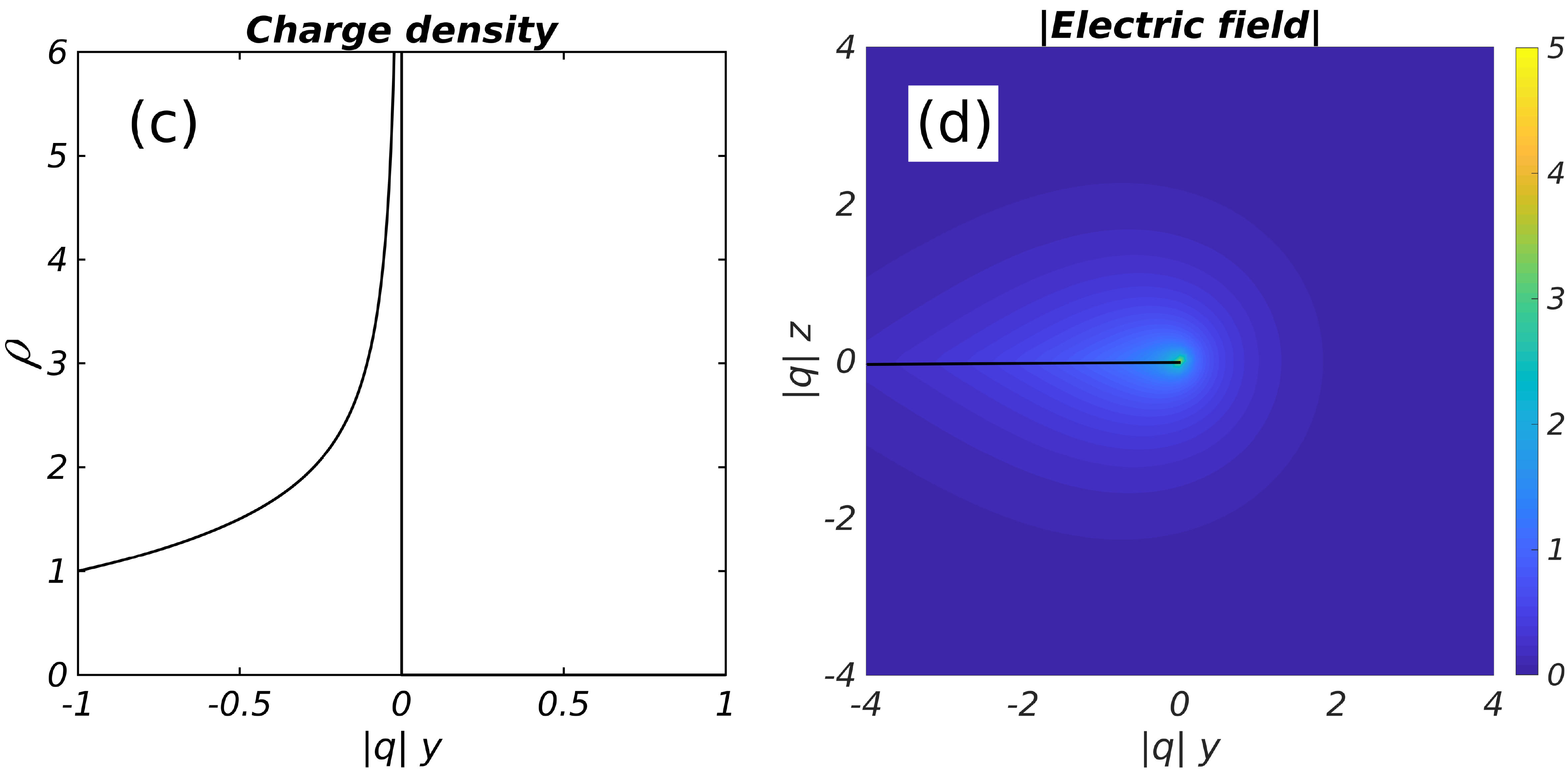}
	\caption{(color online): (a) Charge density of the edge mode in the Fetter approximation  as a function of position, normalized to its value at $|q| y = -1$. (b) Absolute value of the edge-mode near-field
		as a function of position, normalized to its value at $|q| y = -1$.
(c,d) As in corresponding (a,b) panels for the exact solution.		\label{figAbsE}}
\end{figure}

The exact charge profile and field absolute value $|\bm E|$ are plotted in the (c) and (d) panels  of Fig. \ref{figAbsE}, and compare satisfactorily with the approximate results of the upper panels. The  rapid decrease of the field away from the edge is the expected feature of any near field. Although the potential remains finite right at the edge,\cite{Volkov88} the field component in the $yz$ plane diverges as $r^{-1/2} $ when $r\to 0 $,  where $r$ is the distance to the edge.

We are particularly interested in the spin-momentum
locking properties of the near-field that enables non-reciprocity, i.e., directional selectivity in the excitation
of plasmons by means of local sources (dipoles), with appropriately chosen circular (elliptical) polarization.

\begin{figure}
	\includegraphics[width=0.99\columnwidth]{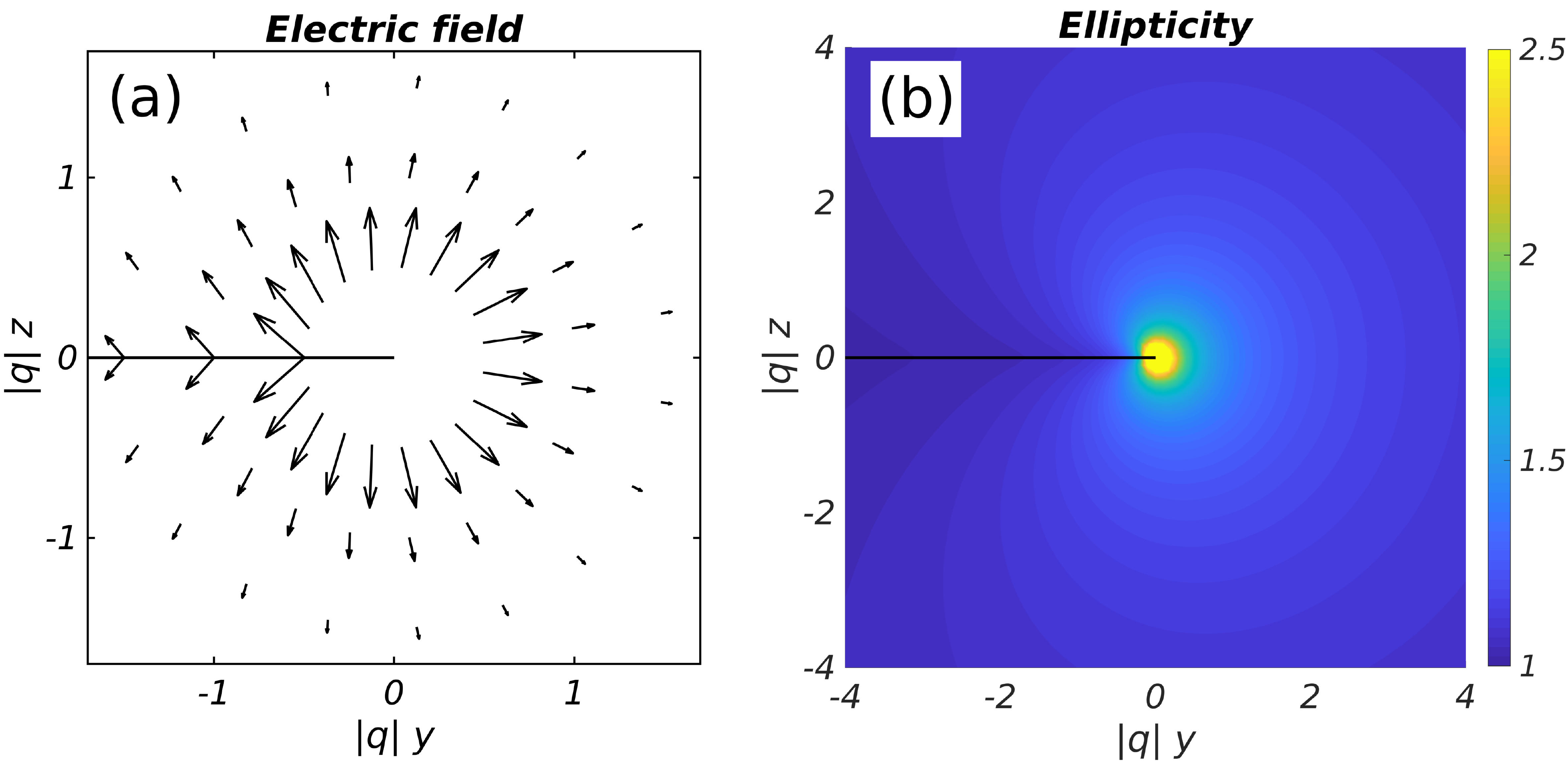}
	\includegraphics[width=0.99\columnwidth]{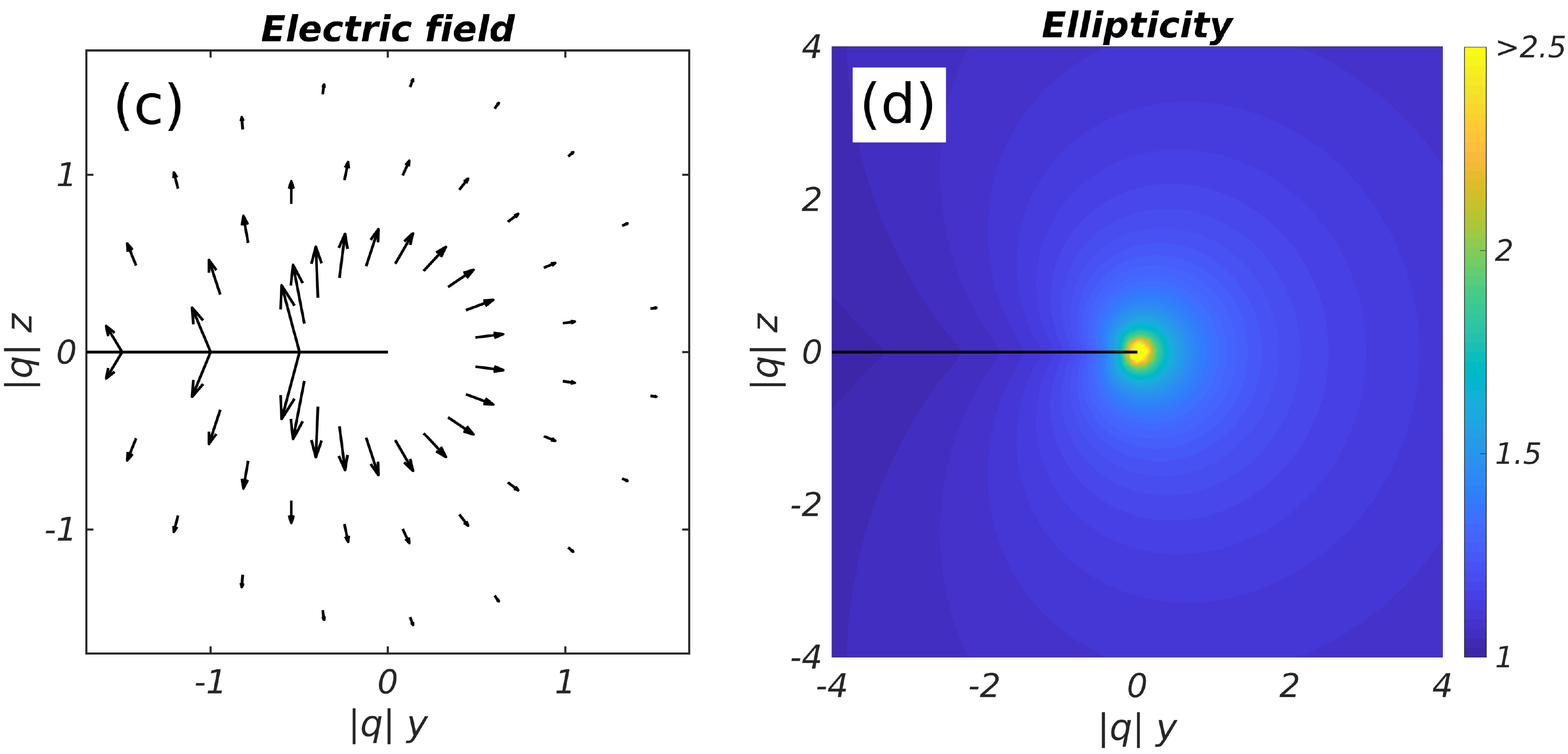}
	\caption{(color online): (a) Field component  in the $yz$ plane created by  the charge profile of the edge mode within the Fetter approximation, for three paths around the edge. (b) Color map of the ellipticity, defined as the ratio of field components $E_{yz}/i E_x $ within the Fetter approximation.
(c,d) As in corresponding (a,b) panels for the exact solution.
\label{figEllip}}
\end{figure}

\begin{figure}
	\includegraphics[width=0.99\columnwidth]{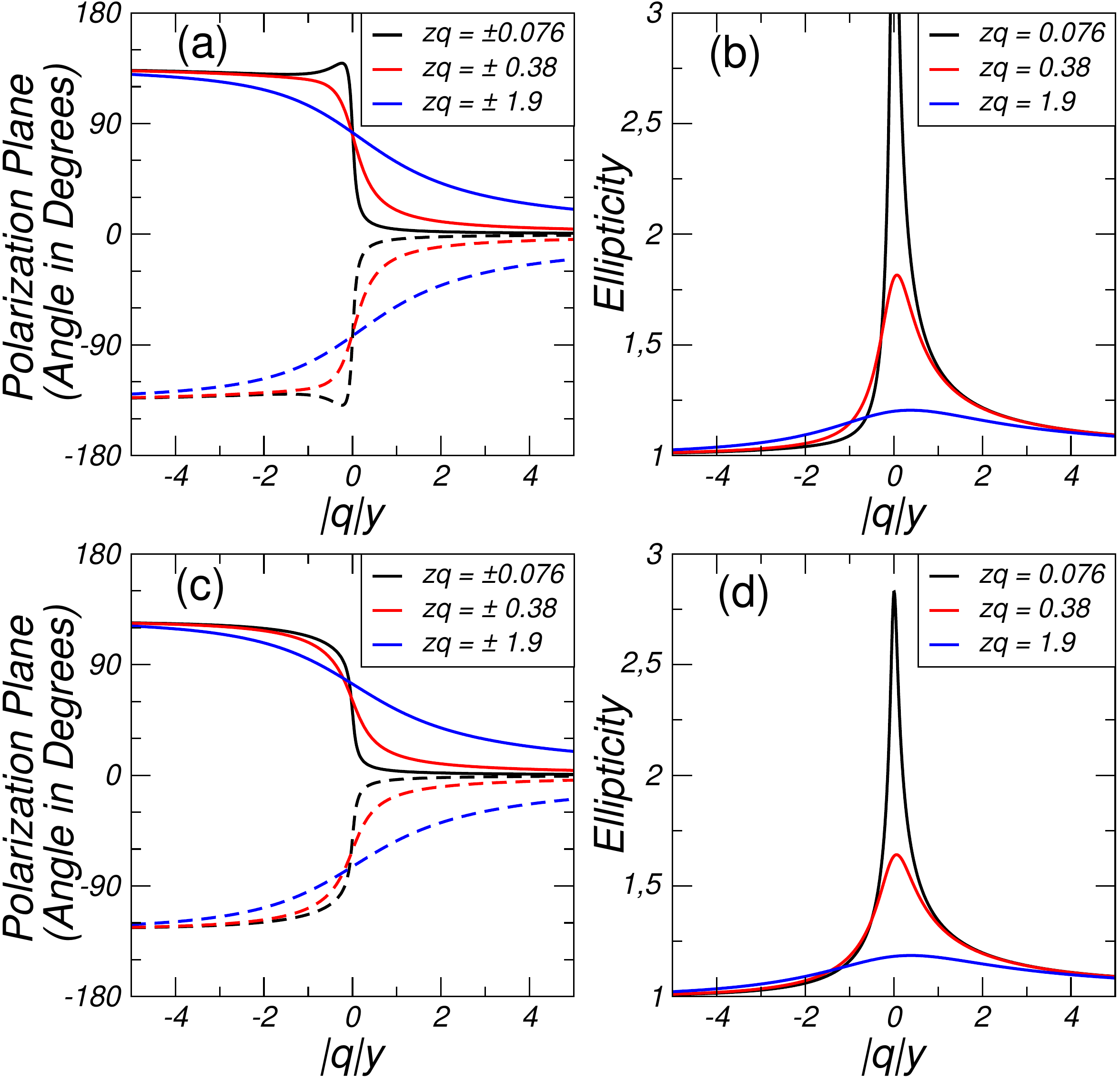}
	\caption{(color online): (a) Polarisation plane as function of $y$ for various values of $z$ within the Fetter approximation. (b) Ellipticity as function of $y$ for various values of $z$ within the Fetter approximation. (c,d) As in corresponding (a,b) panels for the exact solution. The dotted lines on the left hand side correspond to $\sign(z) = -1$ with the same color code as in legend.
		\label{figcombo}}
\end{figure}

In the (c) and (d)  panels of Fig. \ref{figEllip}, we plot the orientation on
the near-field in the $yz$ plane, which together with the
$x$ axis (out of plane in the figure) determines the plane
where the field rotates. This defines the optimal plane
for directional excitation. The ellipticity  or degree of
circular polarization,  $\psi=E_{yz} /iE_x$ is plotted on the right panel
of the same figure. In general, the ellipticity tends to
one (circular polarization) with increasing distance from
the edge. On the contrary, the $yz$ component dominates very close to the edge, where it diverges. Again, the comparison between exact and approximate treatments is satisfactory.

Both features of the exact solution are  summarized in the (c) and (d) panels of Fig. \ref{figcombo}, and again compared with the approximate treatment plotted in the (a) and (b) panels of the same figure. In Fig. \ref{figcombo} (c), we show the rotation of the polarisation plane from
$0^0$ for $y \to +\infty$ to $\pm 125^0$  for $y \to -\infty$ , i.e., the characteristic circling of the polarisation plane around the edge, but
with the asymptotic angle $\theta = \arctan (-\tilde \kappa_z/\tilde \kappa_y) \approx 125^0 $ inside a graphene plane, not far from $135^0 $, the result of the Fetter approximation 
 of Fig.\ref{figcombo} (a). On the right, the ellipticity as
function of $y$ is shown for various distances $z$ with respect to the graphene plane. Far away from the edge,
the ellipticity tends to one, a value that coincides with
that of an infinite line charge within the instantaneous approximation. 

\section{Simulations}
\label{Simulations}
For the near-field of the edge-plasmon and, in appropriate units, the field along the edge ($x$ axis) is always proportional to the phase $i$, whereas the component in the $yz$ plane is real, a quadrature that leads to elliptical polarization. Therefore, an exciting dipole with an elliptical polarization chosen to match the local polarization of an edge-mode with a given $q$, will preferentially excite that mode and, ideally, not the opposite $-q$ mode, in spite of having the same frequency, leading to directional selection\cite{Lodahl17}. 

\begin{figure}
	\centering
	\includegraphics[width = 3in]{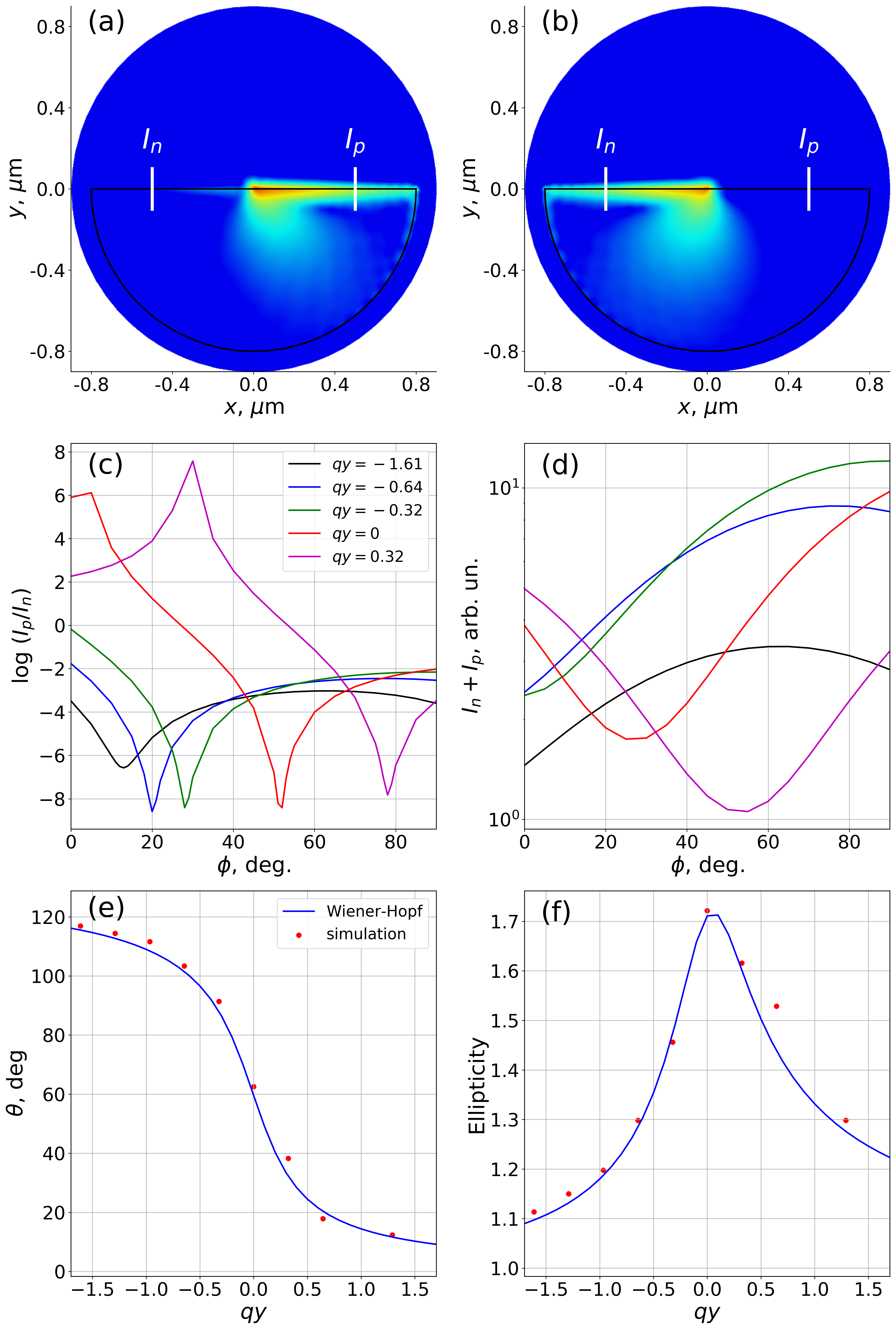}
	\caption{(a,b) Electric field of the edge plasmons excited in a graphene half-disk by an electric dipole, $\mathbf{p} = -i\mathbf{e}_x + 1.4 (\mathbf{e}_y \cos\phi - \mathbf{e}_z \sin\phi)$ A$\cdot$m. The dipole position is $x = y = 0$ nm, $z = 5$ nm ($zq = 0.32$). (a) $\phi = 0^{\circ}$, (b) $\phi = 52^{\circ}$. (c,d) Intensities, $I_a$, $I_b$, of the edge plasmons measured by detectors, indicated by white lines in panels (a,b), as a function of the dipole angle, $\phi$. (e,f) Polarization plane, $\theta$, and ellipticity of the edge mode as a function of $y$. All simulations were performed for the graphene ($\tau = 100$ fs, $\mu = 0.25$ eV) at the frequency $\nu = 40$ THz.   }
	\label{Fig4}
\end{figure}

Here, we will show the results of exact electromagnetic simulations using COMSOL Multiphysics RF Module. We consider a half disk of graphene of radius 800 nm (see Fig. \ref{Fig4}). The graphene's relaxation time and chemical potential are 100 fs and 0.25 eV, respectively. The wavenumber, $q$, of the edge mode, calculated using the dispersion equation Eq. \eqref{DispersionCondition}, is 64.49 $\mu$m$^{-1}$ at the frequency $\nu = 40$ THz. We assume that the edge modes are excited by an electric dipole $\mathbf{p} = -i\mathbf{e}_x + p_0 (\mathbf{e}_y \cos\phi - \mathbf{e}_z \sin\phi)$ A$\cdot$m placed at the point $x = 0$ nm and $z = 5$ nm ($zq = 0.32$). The $y$ coordinate of the dipole is varied as indicated in Fig. \ref{Fig4} (c, d, e, f). 

It can be seen from Figs. \ref{Fig4} (a,b) that we can excite either left or right propagating edge modes by choosing the dipole polarization angle, $\phi$. The angle that allows for the most efficient suppression of the right (left) propagating edge mode increases (decreases) with the increase of the dipole coordinate $y$ (see Fig. \ref{Fig4} (c)). This angle does in general not coincide with the angle which provides the most efficient transfer of the dipole energy to the edge modes (compare Fig. \ref{Fig4} (c) and Fig. \ref{Fig4} (d). Particularly, for the dipole located at $y = 0$, the edge modes carry the highest energy for the dipole polarized at angle $\phi = 90^{\circ}$. On the other hand suppression of the right propagating edge mode requires the dipole polarization to be $\phi = 52^{\circ}$. 

Finally, we want to extract ellipticity and polarization plane of the edge mode from the numerical simulations. In order to do this, we take into account that for the right propagating edge plasmon the mode field structure is $\mathbf{e}_m = - i \mathbf{e}_x + E_0(-\mathbf{e}_y \cos\theta + \mathbf{e}_z \sin\theta)$, where $E_0$ is an ellipticity and $\theta$ is the polarization plane angle. The condition for the mode suppression is $\mathbf{p}\cdot \mathbf{e}_m^* = 0$, which leads to $p_0 E_0 (\cos\phi \cos\theta + \sin\phi\sin\theta) = 1$. By calculating numerically the dipole angle $\phi$ that leads to suppression of the right propagating mode for two different dipole momenta, $p_0 = p_{1,2}$, we can obtain the polarization plane angle and the ellipticity as $\tan\theta = (p_2 \cos\phi_2 - p_1 \cos\phi_1)/(p_1 \sin\phi_1 - p_2\sin\phi_2)$, and $E_0 = 1/(p_1 (\cos\phi_1 \cos\theta + \sin\phi_1 \sin\theta))$. The numerical results are presented in Figs. \ref{Fig4} (e, f) for $p_1 = 1$ and $p_2 = 1.4$. It can be seen that the Wiener-Hopf exact solution provides a very good description of the edge plasmon mode structure in the 2D electron gas as expected, since retardation effects can be neglected. 

\section{Discussion and Conclusion}
In this paper, we have investigated the near field  of edge modes hosted by general two-dimensional materials together with its spin-momentum characteristics. By an analytical and numerical approach, we were able to identify sweet spots to excite edge-modes uniquely in one or two directions. 

We have also extended previous approximate treatments to cover anisotropic and hyperbolic systems, studying the conditions for the existence of edge modes. Furthermore, a detailed assessment of the approximate approach has been carried out, showing it to provide a very good description of the  near field of the edge state when compared with exact results. 

Our analysis is also backed up by exact electromagnetic simulations. By this, we demonstrate the importance of choosing the optimal polarisation plane and position of exciting dipoles. We hope that our results will help to design more efficient plasmonic circuitries. Future studies can further optimise the excitation of unidirectional edge vs. bulk plasmons.  

\section*{Acknowledgements}
This work has been supported by Spain's MINECO under Grant No. FIS2017-82260-P and FIS2015-64886-C5-5-P. TL acknowledges support by the National Science Foundation NSF/EFRI grant (\#EFRI-1741660).

\section*{Appendix A: Derivation of the generalized Fetter solution}
The general incoming vector potential  inside the half-planes $z<0 \; (m=1)$ and $z>0 \; (m=2)$  of the dielectric with $\mu_m$ and $\epsilon_m$ can be written in terms of its Fourier components in $x$-direction as
\begin{align}
\A_m(x,y,z)=\sum_{q_x} e^{iq_xx}\sum_{n=x,y,z}a_{m}^n(q_x,y,z)\e_ne^{-i\omega t}\;,
\end{align}
where $\e_n $  are unit vectors.
Maxwell's equations yield the following wave equation:
\begin{align}
\left(-\epsilon_m\mu_m\frac{\omega^2}{c^2}+q_x^2-\partial_y^2-\partial_z^2\right)a_{m}^n(q_x,y,z)=0
\label{Maxwell}
\end{align}

We will now introduce the Fourier transform in $y$-direction
\begin{align}
a_{m}^n(q_x,y,z)=\int_{-\infty}^{\infty}\frac{dq_y}{2\pi}a_{m}^n(q_x,q_y,z)e^{iq_yy}
\end{align}
With this, we can solve Eq. (\ref{Maxwell}), i.e., $a_{m}^n(q_x,q_y,z)=a_m^n(q_x,q_y)e^{-q_m'|z|}$ where we defined the perpendicular momentum $q_m'=\sqrt{q_x^2+q_y^2-\mu_m\epsilon_m(\omega/c)^2}$ and the velocity of light in vacuum $c^2=1/(\epsilon_0\mu_0)$. The physical fields are obtained as ${\bf E}_m=i\omega\A_m$ and $\mu_m\mu_0{\bf H}_m=\nabla\times\A_m$. The sheet currents are also functions of $y,z$ and shall be expanded in their Fourier series with 
\begin{align}
j^n(q_x,y,z)=\int_{-\infty}^{\infty}\frac{dq_y}{2\pi}j^n(q_x,q_y)e^{iq_yy}\delta(z)
\end{align}

Due to the transverse nature of the vector potential  in the two half-planes, $\nabla\cdot\A_m=0$, we have the following relations:
\begin{align}
iq_xa_1^x+iq_ya_1^y+q_1'a_1^z&=0\\
iq_xa_2^x+iq_ya_2^y-q_2'a_2^z&=0
\end{align}

The boundary condition $\n\times({\bf E}_2-{\bf E}_1)=0$ (with $\n=(0,0,1)$ the normal to the interface) guarantees the continuity of the parallel components of the vector potential, i.e., $a_1^n=a_2^n$ for $n=x,y$. The second boundary condition $\n\times({\bf H}_2-{\bf H}_1)=\j$ yields the two equations with $n=x,y$:
\begin{align}
iq_n\left(\frac{a_2^z}{\mu_2}-\frac{a_1^z}{\mu_1}\right)+q_2'\frac{a_2^n}{\mu_2}+q_1'\frac{a_1^n}{\mu_1}=\mu_0j^n
\end{align}

With the above equations and ($\mu_1=\mu_2=1$), this reduces to 
\begin{align}
\label{BCCurrent}
-q_nq_\np a_2^\np(q_1'+q_2')+((q_1')^2q_2'+(q_2')^2q_1')a_2^n=\mu_0q_1'q_2'j^n\;,
\end{align}
with $n=x,y$ and summation over double indices is implied.

We can then relate the vector potential  to the current densities in the following way:
\begin{align}
\label{BCGauge}
&\left(
\begin{array}{c}
a_m^x(y,z)\\
a_m^y(y,z)
\end{array}\right)=-\frac{\mu_0}{\Omega^2}\int_{-\infty}^\infty\frac{dq_y}{2\pi}e^{iq_yy}e^{-q_m'|z|}\\\notag
&\times\frac{q_1'q_2'}{(q_1')^2q_2'+(q_2')^2q_1'}\left(
\begin{array}{cc}
q_x^2-\Omega^2 &q_xq_y\\
q_xq_y&q_y^2-\Omega^2 
\end{array}\right)
\left(
\begin{array}{c}
j^x\\
j^y
\end{array}\right)\end{align}
with $\Omega^2=\frac{\epsilon_1q_2'+\epsilon_2q_1'}{q_1'+q_2'}\frac{\omega^2}{c^2}$. Notice that in order to get from Eq. (\ref{BCCurrent}) to Eq. (\ref{BCGauge}), one needs to keep the retardation effects because otherwise the determinant of the matrix would vanish.

Let us consider the generating function 
\begin{align}
L_m^{q_x}(y,z)=\frac{1}{2}\int_{-\infty}^\infty \frac{dq_y}{2\pi}e^{iq_yy}e^{-q_m'|z|}\frac{q_1'q_2'}{(q_1')^2q_2'+(q_2')^2q_1'}
\end{align}
which is related to the modified Bessel function of second kind of $z=0$ and in the non-retarded limit $c\to\infty$. We then have 
\begin{align}
\left(
\begin{array}{c}
a_m^x\\
a_m^y
\end{array}\right)=-\frac{\mu_0}{\Omega^2}
\left(
\begin{array}{cc}
q_x^2-\Omega^2&-iq_x\partial_y\\
-iq_x\partial_y&-\partial_y^2-\Omega^2
\end{array}\right)L_m^{q_x}(y,z)
\left(
\begin{array}{c}
j^x\\
j^y
\end{array}\right)\end{align}
with the effective dielectric constant $\epsilon=(\epsilon_1+\epsilon_2)/2$.  

Fourier transforming the current in the above equation, we can thus finally write the relation between the vector potential  and the current in real space:
\begin{align}
a_m^n(q_x,y,z)&=-\frac{\mu_0}{\Omega^2}\left(
\begin{array}{cc}
q_x^2-\Omega^2&-iq_x\partial_y\\
-iq_x\partial_y&-\partial_y^2 -\Omega^2
\end{array}\right)\notag\\
&\times\int_{-\infty}^\infty dy'  L_m^{q_x}(y-y',z)j^n(y')\;.
\end{align}
For a finite current only in the half-space $y<0$, we arrive to Eq. (\ref{EdgeGaugeField}) of the main text where we have slightly changed the notation with $q_x\to \tau q$.

Up to now, all operations have been exact and approximations are necessary to proceed analytically. First, we will neglect retardation effects and second and more crucially, we will approximate the generated function by
\begin{align}
L(y)\approx L_0(y)=\frac{1}{\sqrt{8}}e^{-\sqrt{2}|q_xy|}
\end{align}
since the first three moments of the two functions are identical (after neglecting retardation effects). $L_0(y)$ is now the Green's function with respect to the operator
\begin{align}
\mathcal{O}_0(y)=-\frac{1}{|q_x|}\left(\partial_y^2-2q_x^2\right)\;,
\end{align}
i.e., we have $\mathcal{O}_0(y)L_0(y)=\delta(y)$ and thus
\begin{eqnarray}
\mathcal{O}_0(y)\a(y)=&-\frac{1}{\epsilon_0\epsilon\omega^2}\left(
\begin{array}{cc}
q_x^2&-iq_x\partial_y\\
-iq_x\partial_y&-\partial_y^2 
\end{array}\right)\j(y)\Theta(-y)
\label{Equation}
\end{eqnarray}

The current is generally related to the vector potential  as $j^n=-\chi_{n,\np}a^{\np}$. From Eq. (\ref{Equation}), we see that the $y$-component of the gauge potential must be discontinuous. On the other hand, only the first derivative of the $x$-component of the gauge potential is discontinuous. We can thus make the following ansatz with $\kappa_\pm>0$:
\begin{eqnarray}
\label{ax}
a^x(y)/a_0&=e^{\mp\kappa_\pm y}&\;,\; \pm y>0\\
\label{ay}
a^y(y)/a_0&=\pm i\kappa_\pm/q_xe^{\mp\kappa_\pm y}&\;,\; \pm y>0
\end{eqnarray}
From Eq. (\ref{Equation}), we then get
\begin{align}
\partial_ya^x\Big|_{y=0+}-\partial_ya^x\Big|_{y=0-}&=i\frac{q_x|q_x|}{\epsilon_0\epsilon\omega^2}j^y(0^-)\\
a^y(0^+)-a^y(0^-)&=\frac{|q_x|}{\epsilon_0\epsilon\omega^2}j^y(0^-)\;.
\end{align}
Both expressions yield the same condition, i.e.,
\begin{align}
\label{kappaYApp}
\tilde\kappa_-=\frac{i\tau\tilde\chi_{yx}-\tilde\kappa_+}{1-\tilde\chi_{yy}}
\end{align}
with the dimensionless quantities $\tau=q_x/|q_x|$, $\tilde\kappa_\pm=\kappa_\pm/|q_x|$ and $\tilde\chi_{n,\np}=\frac{|q_x|}{\epsilon_0\epsilon\omega^2}\chi_{n,\np}$.

For $y>0$, the resulting matrix has determinant zero and each component has to be zero. This yields $\kappa_{+}=\sqrt{2}|q_x|$. Therefore, we can write Eq. \ref{kappaYApp} as 
\begin{align}
\label{kappaM}
\tilde\kappa_y\equiv\tilde\kappa_-=\frac{i\tau\tilde\chi_{yx}-\sqrt{2}}{1-\tilde\chi_{yy}}\;,
\end{align}
which is Eq. \ref{kappaY} of the main text.
For $y<0$, we obtain the following condition:
\begin{align}
\label{DispersionConditionApp}
2-\tilde\kappa_y^2= &\tilde\chi_{xx}-\tilde\kappa_y^2\tilde\chi_{yy}-i\tau(\tilde\chi_{xy}+\tilde\chi_{yx})\tilde\kappa_y\;,
\end{align}
which coincides with Eq. \ref{DispersionCondition} of the main text.

\section*{Appendix B: Discussion of the generalized Fetter solution}
 Eqs. \eqref{kappaM} and \eqref{DispersionConditionApp} provide the conditions for the existence of edge modes and can be combined as 
\begin{align}
\label{Combined}
(\tilde\chi_{xy}-i\tau\sqrt2) (\tilde\chi_{yx}+i\tau\sqrt2)=(\tilde\chi_{yy}-1) (\tilde\chi_{xx}-2) \;.
\end{align}
The evanescence of a putative edge mode requires $\operatorname{Re}(\tilde\kappa_y) > 0 $, which for a non-absorbing  system with TRS, where all entries are real, imply $ 1-\tilde\chi_{yy} <0$. Therefore, $\chi_{yy} $ should be at least positive, justifying Eq. \eqref{OrdinaryCondition}. With TRS  where $\tilde\chi_{xy}= \tilde\chi_{yx}$, one can rewrite Eq. \eqref{Combined} as
\begin{equation}
\label{Recombined}
\begin{split}
2 =&  (\tilde\chi_{yy}-1)(\tilde\chi_{xx}-2)-\tilde\chi_{xy}\tilde\chi_{yx}\\
  =& \det (\tilde\chi) + 2 (1-\tilde\chi_{yy}) - \tilde\chi_{xx} \;.
\end{split}
\end{equation}
The right hand side (RHS) of Eq. \eqref{Recombined} vanishes at the edge of the continuum region (see Fig. \ref{figedgedispersion}) and, remembering the definition of tilded quantities: $\tilde\chi=\frac{q}{\epsilon_0\epsilon\omega^2}\chi$, a path in $q$ at constant $\omega$ in the (a) panel of Fig. \ref{figedgedispersion} that goes from the continuum edge to $q \to \infty$ drives the  RHS of Eq. \eqref{Recombined}  from $0$ to $+\infty$ for a  system with $\det (\chi) >0$. This guaranties a solution of Eq. \ref{Recombined} and, therefore, the existence of an edge mode for any ordinary, that is, non-hyperbolic, system with TRS. In the absence of TRS, 
where $\chi_{xy} =  \chi'_{xy} + i \chi_c $, and $\chi_{yx} =  \chi'_{yx} - i \chi_c $, with the real part is still symmetric $\chi'_{xy}=\chi'_{yx} $, 
the previous solution of Eq. \eqref{Combined} simply splits according to the sign of $\tau$, the propagation direction\cite{Volkov88}.
In fact, whenever an edge mode exists with or without TRS, its dispersion relation satisfies
\begin{equation}
\label{FetterDispersionApp}
\omega_e^F(q)= \sqrt{\frac{\chi_{xx} \chi_{yy} - \chi'^{\,2}_{xy} - \chi_c^2}{2 \chi_{yy} +\chi_{xx} -2\sqrt{2} \tau \chi_c} \;  \frac{q}{\epsilon_0\epsilon}  }
,\end{equation}
a result that follows simply from Eq. \eqref{Combined} for the general Drude matrix in non-absorbing systems. 

The previous argument for the existence of edge modes fails for a hyperbolic system, at least under TRS: the RHS of Eq. \eqref{Recombined} cannot be positive and, at the same time, fulfill the requirements $1-\tilde\chi_{yy} <0 $ and $ \det (\tilde\chi) <0$. Therefore, no edge mode detaches from the continuum in hyperbolic systems with TRS.

The analysis for edge modes in the Fetter treatment can be extended to the case where the complementary half-plane, $y > 0 $, is also a material medium with Drude matrix $\chi^{(2)}$, with $\chi^{(1)} $ representing now the response of the original $y < 0 $ half-plane. For the case where both half-planes respect TRS, $\chi^{(1,2)}_{xy} = \chi^{(1,2)}_{yx}$, an edge mode is guaranteed to exist in the following two cases:
\begin{itemize}
\item[i)] medium $(1)$ is elliptic and metallic $(\det(\chi^{(1)})>0, \; \tr(\chi^{(1)})>0) $, and medium $(2)$ is elliptic and capacitive \cite{Bisharat17} $(\det(\chi^{(2)})>0, \; \tr(\chi^{(2)})<0)$.
\item[ii)] medium $(1)$ is elliptic and metallic, and medium $(2)$ is hyperbolic $(\det(\chi^{(2)})<0) $, provided both media share a common region of evanescence in the $(q,\omega) $ plane.
\end{itemize}
Whenever such an edge mode arises, its dispersion within TRS is given by the following rather simple generalization of Eq. \ref{FetterDispersionApp}
\begin{equation}
\label{FetterDispersionGeneral}
\omega_e^F(q)= \sqrt{\frac{\Delta (\det) }{2 \Delta \chi_{yy} + \Delta \chi_{xx}} \;  \frac{q}{\epsilon_0\epsilon}  }
,\end{equation}
with $\Delta (\det) = \det(\chi^{(2)}) -  \det(\chi^{(1)})$, $\Delta \chi_{xx} = \chi^{(2)}_{xx} - \chi^{(1)}_{xx}$, and $\Delta \chi_{yy} = \chi^{(2)}_{yy} - \chi^{(1)}_{yy}$.

\bibliography{Edge_plasmons_bib}

\end{document}